\documentclass[a4paper]{article}

\usepackage[english]{babel}
\usepackage[utf8x]{inputenc}
\usepackage{amsmath}
\usepackage{graphicx}
\usepackage[colorinlistoftodos]{todonotes}

\newtheorem{theorem}{Theorem}
\newtheorem{proposition}{Proposition}

\title{Symmetry reduction and soliton-like solutions for the generalized Korteweg-de Vries equation}

\author{Juan Manuel Conde Mart\'in \& 
David Bl\'azquez-Sanz}

\begin{document}
\maketitle

\begin{abstract}
We analyze the gKdV equation, a generalized version of Korteweg-de Vries with an arbitrary function $f(u)$. In general, for a function $f(u)$ the Lie algebra of symmetries of gKdV is the $2$-dimensional Lie algebra of translations of the plane $xt$. This implies the existence of plane wave solutions. 
Indeed, for some specific values of $f(u)$ the equation gKdV admits a Lie algebra of symmetries of dimension grater than $2$. We compute the similarity reductions corresponding to these exceptional symmetries. We prove that the gKdV equation has soliton-like solutions under some general assumptions, and we find a closed formula for the plane wave solutions, that are of hyperbolic secant type. 
\end{abstract}


\noindent {\it 20MSC:} 35Q53

\section{Introduction}


The generalized Korteveg-de Vries equation (gKdV) is a very natural family of non-linear equations that generalizes the famous  Korteweg-de Vries equation (KdV). This gKdV equation has been studied by several authors using and analytic \cite{Kenig1989} or Lie symmetries \cite{Gungor2004, Bracken2005, Motlatsi2014} approach. It depends on a arbitrary function of $f(u)$. This dependence allows us to obtain some interesting results. The following, is the expression for the gKdV equation:

\begin{equation}\label{gKdV}
u_t = f(u)u_{x} + u_{xxx,}
\end{equation}
We will study its Lie (classical) symmetries and similarity reductions. Our analysis extends some recent results published in \cite{Motlatsi2014}.

\section{Lie symmetries}

For a general exposition of Lie symmetries, the reader may see \cite{Bluman1974,Bluman1989,Olver1993,Stephani1989}. We are interested in Lie symmetries of the gKdV equation \eqref{gKdV}.
We assume that the function $f(u)$ is smooth, somewhere around zero. For different values of $f(u)$ we will obtain different Lie algebras of symmetries. 
Later, we will show that the existence of non trivial symmetries forces to $f(u)$ to be analytic.

From the expression \eqref{gKdV} we easily deduce that the Lie algebra:
\begin{equation}
\left\langle\frac{\partial}{\partial x},\frac{\partial}{\partial t}\right\rangle
\end{equation}
is an algebra of Lie symmetries of \eqref{gKdV}, for any arbitrary function $f(u)$. This is, \eqref{gKdV}  is invariant under translations on the plane $xt$. We are interested in knowing for which specific values of  $f(u)$ this equation possesses a Lie algebra of dimension three or bigger.

\subsection{Lie determining system for gKdV}

We consider a general vector field:
\begin{equation}\label{simetria}
X = \xi(x,t,u) \frac{\partial}{\partial x} + \tau(x,t,u)\frac{\partial}{\partial t} + \eta(x,t,u)\frac{\partial}{\partial u}
\end{equation}
and compute the Lie determining system associated to \eqref{gKdV}. It follows that $X$ is Lie symmetry of \eqref{gKdV} if and only if it satisfies the following system of PDE's:
\begin{eqnarray}
\tau_u &=& 0,\\
\tau_x &=& 0, \\
\xi_u &=& 0, \\
\tau_t &=& 3\xi_x,  \label{eq_tauxi} \\
\xi_{xx} &=& \eta_{xu},   \label{eq_xiXX} \\
\eta_{uu} &=& 0, \\
2f(u)\xi_x + f'(u)\eta + 3\eta_{xxu} + \xi_t &=& \xi_{xxx} \label{eqf_1},\\
\eta_t &=& f(u)\eta_x + \eta_{xxx}. \label{lineal}
\end{eqnarray}

Derivation of \eqref{eq_tauxi} respect to $x$, provides us $\xi_{xx}= 0$. From \eqref{eq_xiXX} we have that $\eta_{xu}=0$. This allows us to simplify equation \eqref{eqf_1}, obtaining:
\begin{equation}
2f(u)\xi_x + f'(u)\eta + \xi_t = 0. \label{eqf_2} \tag{8'}
\end{equation}
Also, derivation of \eqref{lineal} respect to $u$ twice provides:
\begin{equation}\label{eq_eta}
\eta_{tu} = f'(u)\eta_x, \quad 0 = f''(u)\eta_x.
\end{equation}
Hence, we will consider separately the cases $f''(u)=0$ and $f(u)\neq 0$.

\subsection{Case $f''(u) = 0$.}

The case $f''(u) = 0$ is already well known (see for example  \cite{Drazin1989}). In such case $f(u)$ is an affine function:
$f(u) = f_0 + f_1u.$
Whe should consider two qualitatively distinct situations:
\begin{itemize}
\item If $f_1 = 0$ then equation $f(u) = f_0$ is a constant function. Equation \eqref{gKdV} is linear. It admits the infinite dimensional Lie algebra of symmetries:
\begin{equation}
\left\langle\frac{\partial}{\partial x},\, \frac{\partial}{\partial t},\,
-2f_0 t\frac{\partial}{\partial x} + 3t\frac{\partial}{\partial t},\, 
u\frac{\partial}{\partial u}, v(x,t) \frac{\partial}{\partial u}\right\rangle,
\end{equation}
where $v(x,y)$ is any solution of the linear gKdV equation:
\begin{equation}
v_t = f_0v_x + v_{xxx}.
\end{equation}

\item If $f_1 \neq 0 $ the case is equivalent to the classical KdV equation. It admits the fourth dimensional Lie algebra of Lie symmetries:
\begin{equation}
\left\langle\frac{\partial}{\partial x},\, \frac{\partial}{\partial t},\,
2f_0 t\frac{\partial}{\partial x} - 3t\frac{\partial}{\partial t} + 2u \frac{\partial}{\partial u},\, 
f_1t\frac{\partial}{\partial x} + \frac{\partial}{\partial u}\right\rangle.
\end{equation}
\end{itemize}

\subsection{Case $f''(u)\neq 0$.}

Let us now consider the case $f''(u) \neq 0$. First, from equation \eqref{eq_eta}
we obtain $\eta_x = 0$. Applying this into equation \eqref{lineal} yields $\eta_t = 0$. Derivation of \eqref{eqf_2} respect to $t$ provides:
\begin{equation}
\xi_{tt} = -2f(u)\xi_{xt},
\end{equation}
since $\xi$ does not depend on $u$, and we are assuming $f''(u)$ is not identically zero, we can deduce that
\begin{equation}
\xi_{tt}= 0, \quad \xi_{xt} = 0.
\end{equation}
We will now take equations  \eqref{eqf_1}, \eqref{eqf_2}. The solution of the set of all other equations is expressed in terms of six arbitrary constants: $\tau_0$, $a_0$, $a_1$, $b$, $c$, $d$. 
\begin{equation}
\begin{cases}
\tau = \tau_0 + 3bt \\
\xi = a_0 + a_1t + bx \\
\eta = c + du
\end{cases}
\end{equation}
Values of these functions can be replaced in the equation  \eqref{eqf_2}, obtaining an equivalent equation:
\begin{equation}\label{eqf_3}
(c+du)f'(u) + 2bf(u) + a_1 = 0.
\end{equation}
For values of parameters $\tau_0$, $a_0$, $a_1$, $b$, $c$, $d$, when $f(u)$ is a solution fo \eqref{eqf_3} we get a symmetry of \eqref{gKdV}. We note that equation \eqref{eqf_3} does not depend on parameters $\tau_0$ y $a_0$, which correspond to trivial symmetries (translations).

\begin{proposition}\label{proposition1}
Let us assume $f''(u) \neq 0$ in \eqref{gKdV}. A vector field,
$$X = \xi(x,t,u) \frac{\partial}{\partial x} + \tau(x,t,u)\frac{\partial}{\partial t} + \eta(x,t,u)\frac{\partial}{\partial u}$$
is a symmetry of gKdV equation if and only if  
$$
\begin{cases}
\tau = \tau_0 + 3bt \\
\xi = a_0 + a_1t + bx \\
\eta = c + du
\end{cases}
$$
for certain  constant  values $\tau_0$, $a_0$, $a_1$, $b$, $c$, $d$, and furthermore $f(u)$ verifies the following differential equation:
$$
(c+du)f'(u) + 2bf(u) + a_1 = 0.
$$
\end{proposition}

\subsection{Generic and degenerated cases of equation \eqref{eqf_3}}

The general solution of equation \eqref{eqf_3}, if $b$ and $d$ are different of zero, \eqref{eqf_3} is easily calculated by variation of constants. We obtain the following formula:
\begin{equation}\label{solf}
f(u) = \frac{-a_1}{2b} + K(c+du)^{-\frac{2b}{d}},
\end{equation}
where $K$ is an integration constant. There are two degenerated cases.

\begin{itemize}

\item Case $d=0$ and $b\neq 0$. We may assume $c\neq 0$, in other case $f(u)$ would be constant, and we fall into the case of a linear gKdV equation. We obtain this formula for the general solution:
\begin{equation}\label{solf_deg1}
f(u)= -\frac{a_1}{2b} + K \exp\left( -\frac{2b}{c}u \right),
\end{equation}
where  \textit{K} is an integration constant. 

\item Case $b=0$. We may assume $d\neq 0$ for, in other case, we fall in the classical KdV equation. The formula for the general solution is now:
\begin{equation}\label{solf_deg2}
f(u) = \frac{-a_1}{d}\log \left|c+du\right|+ K,
\end{equation}
where $K$ is a constant of integration. 
\end{itemize}

We may state the following:

\begin{proposition}
Let us assume $f''(u) \neq 0$. Equation \eqref{gKdV} possesses a Lie algebra of Lie symmetries of dimension strictly bigger than two if and only if
 $f(u)$ is a function of the form \eqref{solf}, \eqref{solf_deg1} or \eqref{solf_deg2}.
\end{proposition}

An interesting remark is that me may eliminate the parameters $b$, $c$ and $d$ of equation \eqref{eqf_3}. In doing so, we obtain the following single fourth order differential equation with no parameters: 
\begin{equation}\label{equationd2dlog}
\frac{d^2}{du^2}\frac{f(u)'}{f(u)''} = 0.
\end{equation}

\begin{proposition}
Let us assume $f''(u) \neq 0$. Equation \eqref{gKdV} admits a Lie algebra of symmetries of dimension strictly bigger that two if and only if
$$\frac{d^2}{du^2}\frac{f(u)'}{f(u)''} = 0.$$
\end{proposition}

\subsection{General case}

We may eliminate the denominators appearing in equation \eqref{equationd2dlog}, obtaining the following expression:
\begin{equation}
\frac{d^2}{du^2}\frac{f'(u)}{f''(u)} = 
- \frac{f''''(u)f''(u)^2f'(u) + f'''(u)f''(u)^3 
- 2f'''(u)^2f''(u)f'(u)}{f''(u)^4}.
\end{equation}
It is easy to check, that the general solution of this equation contains the case $f''(u) = 0$, and thus the satisfaction of this polynomial equation is the necessary and sufficient condition for equation \eqref{gKdV} to admit a Lie algebra of symmetries of dimension bigger than two. 

\begin{proposition}
Equation \eqref{gKdV} admits a Lie algebra of dimension bigger than two if and only if the function $f(u)$
satisfies the following fourth order differential equation: 
\begin{equation}\label{equationNyS}
f''''(u)f''(u)^2f'(u) + f'''(u)f''(u)^3 
- 2f'''(u)^2f''(u)f'(u) = 0.
\end{equation}
\end{proposition}

In the case in which the equation \eqref{equationd2dlog} is satisfied, the Lie algebra of symmetries is easily computed using Proposition \ref{proposition1}. It always yields a three dimensional Lie algebra of symmetries, extending that of the Lie algebra of translations. We may state our final result on the symmetries. This result slightly clarifies that of \cite{Motlatsi2014}, that allowed some overlapping of cases in their classification table. 

\begin{theorem}
Let us consider equation \eqref{gKdV} where $f(u)$ is a function defined in some neighborhood of the origin. It falls in exactly one of the following six cases (a), (b.1), (b.2), (b.3.i), (b.3.ii) or (b.3.iii).
\begin{itemize}
\item[(a)] The function $f(u)$ does not satisfy equation \eqref{equationNyS}. The Lie algebra of symmetries is the two dimensional Lie algebra of infinitesimal translations of the plane $xt$,
$$\left\langle\frac{\partial}{\partial x},\frac{\partial}{\partial t}\right\rangle.$$
\item[(b)] The function $f(u)$ satisfies equation \eqref{equationNyS}. We have the following subcases:
\begin{itemize}
\item[(b.1)] $f'(u) = 0$, in such case, $f(u)$ is a constant function: $f(u) = f_0$. Equation \eqref{gKdV} is a linear PDE and it admits the infinite dimensional Lie algebra of symmetries:
$$\left\langle\frac{\partial}{\partial x},\, \frac{\partial}{\partial t},\,
-2f_0 t\frac{\partial}{\partial x} + 3t\frac{\partial}{\partial t},\, 
u\frac{\partial}{\partial u}, v(x,t) \frac{\partial}{\partial u}\right\rangle,$$
where $v(x,y)$ is any solution of the linear gKdV equation:
$$v_t = f_0v_x + v_{xxx}.$$
\item[(b.2)] $f'(u) \neq 0$ and $f''(u) = 0$, in such case $f(u) = f_0 + f_1u$ is an affine function. Equation  \eqref{gKdV} is a classical KdV equation and it admits the following fourth dimensional Lie algebra of symmetries:
$$\left\langle\frac{\partial}{\partial x},\, \frac{\partial}{\partial t},\,
2f_0 t\frac{\partial}{\partial x} - 3t\frac{\partial}{\partial t} + 2u \frac{\partial}{\partial u},\, 
f_1t\frac{\partial}{\partial x} + \frac{\partial}{\partial u}\right\rangle. $$
\item[(b.3)] $f''(u) \neq 0$ and $\frac{d^2}{du^2}\frac{f'(u)}{f''(u)} = 0$.

\begin{itemize}
\item[(b.3.i)] $f$ is of the form 
$f(u) = f_0 + \lambda(u-u_0)^\alpha$
with $\lambda\neq 0$, $\alpha \neq 0,1$. Equation \eqref{gKdV} admits the following three dimensional Lie algebra of symmetries:
$$\left\langle \frac{\partial}{\partial x}, \frac{\partial}{\partial t}, (2f_0t-x)\frac{\partial}{\partial x}- 3t\frac{\partial}{\partial t}+ \frac{2(u-u_0)}{\alpha}\frac{\partial}{\partial u} \right\rangle.$$
\item[(b.3.ii)] $f$ is of the form 
$f(u) = f_0 + \lambda e^{\alpha u}$
with $\lambda\neq 0$, $\alpha \neq 0,1$. Equation \eqref{gKdV} admits the following three dimensional Lie algebra of symmetries:
$$\left\langle \frac{\partial}{\partial x}, \frac{\partial}{\partial t}, (2f_0t-x)\frac{\partial}{\partial x}- 3t\frac{\partial}{\partial t}+ \frac{2}{\alpha}\frac{\partial}{\partial u} \right\rangle.$$
\item[(b.3.iii)] $f$ is of the form $f(u) = f_0 + \alpha \log|u-u_0|$, with $\alpha\neq 0$. In this case \eqref{gKdV} admits the following three dimensional Lie algebra of symmetries:
$$\left\langle \frac{\partial}{\partial x}, \frac{\partial}{\partial t}, -\alpha t \frac{\partial}{\partial x}+(u-u_0)\frac{\partial}{\partial u} \right\rangle.$$
\end{itemize}
\end{itemize}
\end{itemize}
\end{theorem}

Let us note that out result is of algebraic nature. We do not state that if a gKdV equation admits an additional symmetry, then it is equivalent to some of the above cases, but indeed, it belongs to one of the above cases. By an affine change of variables in $u$, parameters $f_0$, $f_1$ from case $(b.2)$, $\lambda$, $u_0$ from case $(b.3.i)$, $\lambda$, $\alpha$ from case $(b.3.ii)$  and $f_0$, $u_0$ from case $(b.3.iii)$ are eliminated. Then, we can state that if a gKdV equation \eqref{gKdV} admits a Lie algebra of symmetries of dimension strictly bigger than two then it is equivalent to one of the following cases:

\begin{center}
  \begin{tabular}{ | c | c |}
    \hline
    $f(u)$ & Lie algebra of symmetries \\ \hline 
    & \\
    $1$   & 
    $\left\langle\frac{\partial}{\partial x},\, \frac{\partial}{\partial t},\,
-2t\frac{\partial}{\partial x} + 3t\frac{\partial}{\partial t},\, 
u\frac{\partial}{\partial u}, v(x,t) \frac{\partial}{\partial u}\right\rangle$, $v$ satisfies \eqref{lineal}  \\ 
    $u$   & 
    $\left\langle\frac{\partial}{\partial x},\, \frac{\partial}{\partial t},\,
 - 3t\frac{\partial}{\partial t} + 2u \frac{\partial}{\partial u},\, 
t\frac{\partial}{\partial x} + \frac{\partial}{\partial u}\right\rangle$ \\
    $f_0 + u^\alpha$,  \,\, $\alpha \neq 0, 1$ & 
    $\left\langle \frac{\partial}{\partial x}, \frac{\partial}{\partial t}, (2f_0t-x)\frac{\partial}{\partial x}- 3t\frac{\partial}{\partial t}+ \frac{2u}{\alpha}\frac{\partial}{\partial u} \right\rangle.$  \\
    $f_0 + e^u$, &
    $\left\langle \frac{\partial}{\partial x}, \frac{\partial}{\partial t}, -x\frac{\partial}{\partial x}- 3t\frac{\partial}{\partial t}+ \frac{2}{\alpha}\frac{\partial}{\partial u} \right\rangle$\\
    $\alpha \log(u-1)$, \,\,  $\alpha \neq 0$ & $\left\langle \frac{\partial}{\partial x}, \frac{\partial}{\partial t}, -\alpha t \frac{\partial}{\partial x}+(u-1)\frac{\partial}{\partial u} \right\rangle$\\ & \\
    \hline
  \end{tabular}
\end{center}

This completes the table appearing in \cite{Olver1996, Gungor2004}, that seems to miss the logarithmic case, and clarifies that appearing in \cite{Motlatsi2014}, in which some equivalent equations are redundantly listed.




\section{Similarity reductions}

\subsection{Plane wave solutions in the general case}

For any value of $f(u)$ equation \eqref{gKdV} is invariant by translations on the plane $xt$. The similarity reductions obtained by this kind of symmetries are the ordinary differential equations satisfied by plane wave solutions. Thus, let us look for 
solutions of \eqref{gKdV} of the form:
\begin{equation}
u(x,t) = w(z) , \quad z = x + ct.
\end{equation}
By direct substitution, we obtain the reduced equation for $w$:
\begin{equation}
-c\frac{dw}{dz} + f(w)\frac{dw}{dz} + \frac{d^3 w}{dz^3}  = 0.
\end{equation}
Let us take $F(w) = \int_0^w f(\xi)d\xi$. Then, we may integrate and obtain the differential equation:
\begin{equation}
\frac{d^2 w}{dz^2} + F(w) - cw = k,
\end{equation}
where $k$ is a constant of integration. The dynamics of such equation, is described by a classical hamiltonian system in the plane $(w,\dot w)$, where $\dot w$ represents the
derivative of $w$ with respect to $z$, of energy $H(w,\dot w)$:
\begin{equation}\label{hamiltonian}
H(w,\dot w) = \frac{\dot w^2}{2} + V(w),
\end{equation}
where the potential is given by:
\begin{equation}\label{potential}
V(w) = -\frac{cw^2}{2} + kw + \int_0^w F(\xi)d\xi.
\end{equation}
Finally, by conservation of energy, the equation is reduced to the first order ordinary differential equation:
\begin{equation}\label{1st_order_pw}
\left(\frac{dw}{dz}\right)^2 = 2E + cw^2 - 2kw - 2\int_0^w F(\xi)d\xi,
\end{equation}
that depends on the arbitrary constants $c$, $k$ and $E$.

\subsection{Reductions of \eqref{gKdV}, for  $f(u) = f_0 + (u - u_0)^\alpha$}

Let us consider the gKdV equation \eqref{gKdV}, with $f = f_0 + (u-u_0)^\alpha$. A symmetry of this equation is a vector field of the form:
\begin{equation}
X = \xi \frac{\partial}{\partial x} + \tau   \frac{\partial}{\partial y} + \eta  \frac{\partial}{\partial u}, \quad\mbox{with}\quad  \begin{cases}
\xi =(2f_0t-x)c_1+c_3 \\
\tau = -3c_1t+c_2 \\
\eta =2 \frac{u-u_0}{\alpha} c_1
\end{cases}
\end{equation}
Therefore, characteristics equation takes the form:
\begin{equation}\label{car1}
{\frac{{\it dx}}{\left(2\,f_{0}\,t-x\right)c_{1}+c_{3}}}={\frac{{\it dt}}{-3\,c_{1}\,t+c_{2}}}={\frac{{\it du}\,\alpha}{2c_{1}\,\left(u-u_{0}\right)}}.
\end{equation}
  The case $c_1 = 0$ corresponds to the plane wave solutions, that have been already studied in the general case. Let us assume $c_1 \neq 0$. Any pair of symmetries with $c_1 \neq 0$ are conjugated by a translation. Hence, we can set without loss of generality $c_{2}=0$, $c_{3}=0$. The characteristics equation is now:
\begin{equation}\label{car1a}
{\frac{{\it dx}}{2\,f_{0}\,t-x}}=-\frac{1}{3}\,{\frac{{\it dt}}{t}}={\frac{{\it du}\,\alpha}{2\,(u-u_{0})}}.
   \end{equation}
Considering $ f_{0}\neq 0$. Then, from above equation, we obtain the variables for the reduction:
\begin{equation}
u=u_{0}+w\left(z\right){t}^{\frac{-2}{3\alpha}}  ,\,z={\frac{f_{0}\,t+x}{\sqrt[3]{t}}}.
\end{equation}
 Substituting in  \eqref{gKdV},  we obtain the reduced equation;  
\begin{equation}\label{red1}
3\alpha w^\alpha \frac{dw}{dz} + z \alpha \frac{dw}{dz} + 3\alpha \frac{d^3 w}{dz^3} + 2w = 0.
\end{equation}
 This equation does not admit, in general, a quadrature as it happened with the equation for plane wave solutions. However, in  the case  $\alpha=2$, which corresponds to a version of the modified Korteweg-de Vries equation (widely known as mKdV), we can easily integrate, to obtain the following second order equation:
\begin{equation}
 3 \frac{d^2w}{dz^2} + zw + w^3 = k,
\end{equation}
 where $k$ is a constant of integration. Is not difficult to check that, up to some complex linear change of variable in $w$ and $z$, this equation is equivalent to the well known second Painlev\'e equation. 
 
 We have assumed so far, $f_0 \neq 0$. However, in the case $f_0=0 $, characteristics equations converts into:
 \begin{equation}\label{car1aa}
{\frac{{\it dx}}{-x}}=-\frac{1}{3}\,{\frac{{\it dt}}{t}}={\frac{{\alpha\it du}}{2\,(u-u_{0})}}.
\end{equation}
If we solve this equation, we have exactly the same reduction when is  $f_0\neq 0$, obtaining the same reduced equation \eqref{red1}.
 
\subsection{Reductions of \eqref{gKdV}, for $f(u) =f_{0}+\lambda\,{{\rm e}^{\alpha\,u}}$}


Let us consider the gKdV equation \eqref{gKdV} with $f(u) = f_{0}+\lambda\,{{\rm e}^{\alpha\,u}}$. A symmetry of this equation is a vector field of the form:
\begin{equation}
X = \xi \frac{\partial}{\partial x} + \tau   \frac{\partial}{\partial y} + \eta  \frac{\partial}{\partial u}, \quad \mbox{with}\quad 
\begin{cases}
\xi =(2f_0t-x)\ c_1+c_3 \\
\tau = -3 c_1 t+c_2 \\
\eta =2 \frac{c_1}{\alpha}
\end{cases}
\end{equation}
Characteristics equation is now:
 \begin{equation}\label{car3}
{\frac{{\it dx}}{\left(2\,f_{0}\,t-x\right)c_{1}+c_{3}}}={\frac{{\it dt}}{-3\,c_{1}\,t+c_{2}}}={\frac{{\it du}\,\alpha}{2c_{1}}}.
\end{equation}

As in the previous section, the case $c_1 = 0$ corresponds to the plane wave solutions, that have already been discussed in the general case. Let us consider $c_1 \neq 0$. Again, any pair of symmetries with $c_1 \neq 0$ are conjugated by a translation, so we can assume, without loss of generality, $c_{2}=0$  and $c_{3}=0$. Characteristics equation takes the form:
\begin{equation}\label{car1a}
{\frac{{\it dx}}{2\,f_{0}\,t-x}}=-\frac{1}{3}\,{\frac{{\it dt}}{t}}={\frac{{\it du}\,\alpha}{2}}.
   \end{equation}
Let us assume $ f_{0}\neq 0$. From above equation, we compute the variables for the reduction:
 \begin{equation}
  u=-\frac{2}{3}\,{\frac{\ln\left(t\right)}{\alpha}}+w\left(z\right),\,z={\frac{f_{0}\,t+x}{\sqrt[3]{t}}}.
\end{equation}
By substituting into \eqref{gKdV}, we obtain the reduced equation:
\begin{equation}\label{red2}
3\alpha \frac{d^3w}{dz^3}  +\alpha \left(3\lambda e^{\alpha w} + z \right) \frac{dw}{dz} + 2 = 0.
\end{equation}
Finally, the case with $f_0=0 $ , provides exactly the same reduction that the case $f_0\neq 0$, therefore, we have again \eqref{red2} itself.
 
\subsection{Reductions of \eqref{gKdV}, for  $ f(u)=f_0 + \alpha \log|u-u_0|$.}

Let us consider the gKdV equation \eqref{gKdV}, with $f(u) = f_0 + \alpha \log|u-u_0|$. A symmetry of this equation is a vector field of the form:
 \begin{equation}
X = \xi \frac{\partial}{\partial x} + \tau   \frac{\partial}{\partial y} + \eta  \frac{\partial}{\partial u}, \quad \mbox{with}\quad 
\begin{cases}
\xi =-c_2 t+c_3 \\
\tau = c_1 \\
\eta =\frac{c_2}{\alpha} (u-u_0)
\end{cases}
\end{equation}
In what follows, we will suppose $u-u_0>0$, so that, the function $f(u)$ is defined in a neighborhood of $0$. 
Characteristic equation takes the form:
\begin{equation}\label{car3}
\frac{dx}{-c_2 t+c_3}=\frac{dt}{c_1}=\frac{du}{\frac{c_2}{\alpha} (u-u_0)}.
\end{equation}
Let us remember, that our equation is invariant under temporal translations, then, we can take $c_3 =0$. In that case, we can choose $c_2 =1$, then characteristics equation is:
\begin{equation}
-{\frac{{\it dx}}{t}}=\frac{dt}{c_1}={\frac{\alpha\,{\it du}}{u-u_{0}}}.
\end{equation}
It is easy to deduce
\begin{equation}
u=u_{0}+w\left(z\right){{\rm exp}\left(\frac{t}{c_{1}\alpha}\right)},z=x+\frac{1}{2}{\frac{{t}^{2}}{c_{1}}},
\end{equation}
which substitution into  \eqref{gKdV}, gives us
\begin{equation}
c_1\alpha\frac{d^3w}{dz^3} + c_1 \alpha(\alpha\ln(w)+f_0)\frac{dw}{dz} - w = 0.
\end{equation}

The logarithm function can be eliminated, with the following change of the dependent variable, 
 \begin{equation}
 w\left(z\right)={{\rm exp}({y\left(z\right)})},
    \end{equation}
it converts into
\begin{equation}
 \frac{d^3y}{dz^3}  + \left(\frac{dy}{dz}\right)^3 + (\alpha y + f_0) \frac{dy}{dz}
 + 3\frac{dy}{dz} \frac{dy^2}{dz^2}
 = \frac{1}{c_1\alpha}.
\end{equation}
The order of above equation can be reduced by one. We take:
\begin{equation}
\frac{d}{dz}y(z) =p\left(y\left(z\right)\right),\,\, y\left(z\right)=\theta,
 \end{equation}
to obtain a second order equation for $p(\theta)$:
\begin{equation}
  p^2\frac{d^2p}{d\theta^2} + 3p^2\frac{dp}{d\theta} +
  p^3 + (\alpha \theta + f_0)p   = \frac{1}{c_1\alpha}.
\end{equation}

\section{Solitary waves }

\subsection{Plane wave solutions with a single critical point and exponential decay at infinity}
 
Here we will look for plane wave solutions of \eqref{gKdV} of the form:
\begin{equation}
u(x,t) = w(z), \quad z = x+ct,
\end{equation}
that are exponentially small at $z = \pm \infty$. For this analysis we rely on the hamiltonian formulation \eqref{hamiltonian} of the equation for plane wave solutions of \eqref{gKdV}. We only assume, $f(u)$ to be continuous in a neighborhood of $0$. 

A solution with exponential decay at $z = \pm \infty$, should be an homoclinic curve of the Hamiltonian system \eqref{hamiltonian} associated to a critical point at the origin $(0,0)$ in the plane $(w,\dot w)$. That means, a solution $(w(z),\dot w(z))$ of the hamultonian system \eqref{hamiltonian} such that (see, for instance, \cite{Arnold1978} \S 12.2):
\begin{equation}
\lim_{z\to \pm\infty}(w(z),\dot w(z))) = (0,0).
\end{equation}
In order to have critical point of $(0,0)$, we need to take $k = 0$ in \eqref{potential}. If we assume that $f$ is continuous near $0$, then we have that the function:
\begin{equation}
h(w) = \frac{1}{w^2}\int_0^w \int_0^\eta f(\xi)d\xi d\eta,
\end{equation}

is continuous and takes the value $h(0) = \frac{f(0)}{2}$. The potential is thus written as:
\begin{equation}\label{potential2}
V(w) = \left(h(w) - \frac{c}{2} \right)w^2.
\end{equation}
In such case, we may use the energy invariant to solve the equations of movement, by means of the inversion of an integral obtaining:
\begin{equation}\label{integral_inversion}
z = \int_{w_0}^{w(z)} \frac{dw}{w\sqrt{c - 2h(w)}}.
\end{equation}
In the case in which $w_0$ is in a potential well, between $0$ and a single zero of $V(w)$, the function $w(z)$ in formula \eqref{integral_inversion}, corresponds to an homoclinic curve of the Hamiltonian system \eqref{hamiltonian}, it has exponential decay to $0$ at $z = \pm \infty$. 

\begin{proposition}
Let us assume that $F(z)$ is positive in the interval $(0,\lambda)$. Then, for each 
$w_0\in (0,\lambda)$ the formula: 
$$z = \int_{w_0}^{w(z)} \frac{dw}{w\sqrt{2h(w_0) - 2h(w)}}$$
gives a plane wave solution
$$u(x,t) = w(z) \quad z = x + h(w_0)t,$$
where the function $w(z)$ takes a maximum value of $w_0$ at $z= 0$, has no other critical points, and decays exponentially to zero as $z\to \pm\infty$. 
\end{proposition}

\subsection{Hyperbolic secant soliton-type solutions}

It is well known, that existence of solitons described by hyperbolic functions
is typical of completely integrable equations, see for instance \cite{Zabusky1965, Drazin1989}; this is, solutions of hyperbolic secant type. We have found the existence of traveling wave solutions for \eqref{gKdV}. In the particular case of $f(u) = f_0 + (u-u_0)^\alpha$, these solitary waves are described by hyperbolic secants, we found them by a direct search. 
 
 \begin{theorem}\label{th_soliton}
Let us assume $\alpha \neq 0, -1, -2$ and $f(u) = f_0 + (u-u_0)^\alpha$. 
The function:
\begin{equation}
u(x,t) = c + a {\rm sech}^\beta(A(x-c_3t)+b)
\end{equation}
is a solution of \eqref{gKdV} 
if and only if $A$, $\beta$ satisfy the following system of algebraic equations:
\begin{equation}
\begin{cases}
A^{2}(\beta+1)(\beta+2)-a^{2/\beta}=0
 \\
f_{0}+c_{3}+A^{2}\beta^{2}=0
  \\
\alpha=\frac{2}{\beta}
   \\
   u_{0}=c
 \end{cases}
\end{equation}
\end{theorem}

A particular case of this result can be found in \cite{Smyth1995}. Let us discuss some remarks on Theorem \ref{th_soliton}:

\begin{itemize}
\item[(a)] This equation has this kind of solutions  $\forall\beta\neq -1,-2
$. However, it can only be considered as a physically acceptable soliton for $\beta>0 $.

\item[(b)] If we suppose that $\beta>0$.  Then, for $\alpha $   large, $\beta $   is small and then, solitons are wide. Otherwise, for $\alpha $   small, $\beta $ is large, and then, solitons are narrow. Is important to note, that all these solitons travel at the same speed.

\item [(c)] 
These  results are completely valid for $\alpha = 1$, then, $\beta =2$, this situation corresponds to equation KdV itself.
\end{itemize} 
 
Let us see some examples of hyperbolic secant soliton-type solutions:

\begin{itemize}
\item[(1)] If  $\alpha = 1$, then, $\beta =2$, therefore $a=12A^2$ , $ u_{0}=c$  and $f_{0}+c_{3}+4A^2=0$. Soliton solutions writes as
\begin{equation}
u(x,t) = -u_0+a {\rm sech}^2 (A(x-c_3t)+b),
\end{equation}
which is the well known $1$-soliton solution for the classical KdV equation.

\item[(2)] If  $\alpha = 2$, then, $\beta =1$, therefore $a^2=6A$ , $ u_{0}=c$  and $f_{0}+c_{3}+A^2=0$. Soliton solutions writes as
\begin{equation}
u(x,t) = -u_0+a{\rm sech} (A(x-c_3t)+b),
\end{equation}
that is the well known $1$-soliton solution for mKdV.
   
 \item[(3)]  If  $\alpha = \sqrt[]{2}$, then, $\beta =\sqrt[]{2}$, therefore $A^2(4+3\sqrt[]{2})=a^{\sqrt[]{2}}$, $ u_{0}=c$ and $(f_{0}+c_{3})+2A^2=0$.
Then, this new soliton solution writes as
\begin{equation}
u(x,t) = -u_0+a {\rm sech}^{\sqrt[]{2}}(A(x-c_3t)+b), 
\end{equation}
which corresponds to a soliton of width, between KdV and mKdV ones, but traveling at the same speed.
\end{itemize}

\section{Conclusions}

We have studied a version of the Korteweg de Vries equation, known as gKdV. This equation has been previously studied by various authors, however, we have extended previous analysis and obtained new results. We have clarified a result of the Lie algebras, in addiction, we have found a few differential equations satisfied by the arbitrary function $f(u)$. 

We have proved, that this equations possesses, in general, plane wave solutions with a single critical point with an exponential decay at infinity. We have studied completely all of the similarity reductions (coming from Lie classical symmetries) for this gKdV equation. In particular, we have found explicit expressions for solitons, for the case of $f(u) = f_0 + (u-u_0)^\alpha$, where $\alpha$ is a real and strictly positive number. All of these solitons travel at the same speed, but have different width, depending on $\alpha$.

\subsection*{Acknowledgements}

DBS acknowledges the support of Universidad Nacional de Colombia through project HERMES 11368. JMCM acknowledges the support of Universidad San Francisco de Quito.
We also want to express our gratitude to prof. F. G\"ung\"or, who kindly pointed us out some mistakes in the first version of the manuscript and let us know about some useful references.

\bibliographystyle{plain}
\bibliography{biblio.bib}

{\sc
\noindent Juan Manuel Conde Mart\'in \\
Universidad San Francisco de Quito, Ecuador}  \\
{\tt jconde@usfq.edu.ec} \\

\medskip
{\sc
\noindent David Bl\'azquez-Sanz \\
Universidad Nacional de Colombia - Sede Medell\'in, Colombia} \\
{\tt dblazquezs@unal.edu.co}

\end{document}